\documentclass[aip, apm, preprint, amsmath, graphicx]{revtex4-1}

\usepackage{graphics}
\usepackage{graphicx}
\usepackage{bm}
\usepackage{color}

\begin{document}


\title{Thermoelectric performance of spin Seebeck effect in Fe$_3$O$_4$/Pt-based thin film heterostructures}



\author{R. Ramos}
\email[]{ramosr@imr.tohoku.ac.jp}
\affiliation{WPI Advanced Institute for Materials Research, Tohoku University, Sendai 980-8577, Japan}
\affiliation{Spin Quantum Rectification Project, ERATO, Japan Science and Technology Agency, Sendai 980-8577, Japan}

\author{A. Anad\'{o}n}
\affiliation{Instituto de Nanociencia de Arag\'{o}n, Universidad de Zaragoza, E-50018 Zaragoza, Spain}
\affiliation{Departamento de F\'{i}sica de la Materia Condensada, Universidad de Zaragoza, E-50009 Zaragoza, Spain}

\author{I. Lucas}
\affiliation{Instituto de Nanociencia de Arag\'{o}n, Universidad de Zaragoza, E-50018 Zaragoza, Spain}
\affiliation{Departamento de F\'{i}sica de la Materia Condensada, Universidad de Zaragoza, E-50009 Zaragoza, Spain}

\author{K. Uchida}
\affiliation{Institute for Materials Research, Tohoku University, Sendai 980-8577, Japan}
\affiliation{PRESTO, Japan Science and Technology Agency, Saitama 332-0012, Japan}

\author{P. A. Algarabel}
\affiliation{Departamento de F\'{i}sica de la Materia Condensada, Universidad de Zaragoza, E-50009 Zaragoza, Spain}
\affiliation{Instituto de Ciencia de Materiales de Arag\'{o}n, Universidad de Zaragoza and Consejo Superior de Investigaciones Cient\'{i}ficas, 50009 Zaragoza, Spain}

\author{L. Morell\'{o}n}
\affiliation{Instituto de Nanociencia de Arag\'{o}n, Universidad de Zaragoza, E-50018 Zaragoza, Spain}
\affiliation{Departamento de F\'{i}sica de la Materia Condensada, Universidad de Zaragoza, E-50009 Zaragoza, Spain}

\author{M. H. Aguirre}
\affiliation{Instituto de Nanociencia de Arag\'{o}n, Universidad de Zaragoza, E-50018 Zaragoza, Spain}
\affiliation{Departamento de F\'{i}sica de la Materia Condensada, Universidad de Zaragoza, E-50009 Zaragoza, Spain}
\affiliation{Laboratorio de Microscop\'{i}as Avanzadas, Universidad de Zaragoza, E-50018 Zaragoza, Spain}

\author{E. Saitoh}
\affiliation{WPI Advanced Institute for Materials Research, Tohoku University, Sendai 980-8577, Japan}
\affiliation{Spin Quantum Rectification Project, ERATO, Japan Science and Technology Agency, Sendai 980-8577, Japan}
\affiliation{Institute for Materials Research, Tohoku University, Sendai 980-8577, Japan}
\affiliation{Advanced Science Research Center, Japan Atomic Energy Agency, Tokai 319-1195, Japan}

\author{M. R. Ibarra}
\affiliation{Instituto de Nanociencia de Arag\'{o}n, Universidad de Zaragoza, E-50018 Zaragoza, Spain}
\affiliation{Departamento de F\'{i}sica de la Materia Condensada, Universidad de Zaragoza, E-50009 Zaragoza, Spain}
\affiliation{Laboratorio de Microscop\'{i}as Avanzadas, Universidad de Zaragoza, E-50018 Zaragoza, Spain}


\date{\today}

\begin{abstract}
We report a systematic study on the thermoelectric performance of spin Seebeck devices based on Fe$_3$O$_4$/Pt junction systems. We explore two types of device geometries: a spin Hall thermopile and spin Seebeck multilayer structures. The spin Hall thermopile increases the sensitivity of the spin Seebeck effect, while the increase in the sample internal resistance has a detrimental effect on the output power. We found that the spin Seebeck multilayers can overcome this limitation since the multilayers exhibit the enhancement of the thermoelectric voltage and the reduction of the internal resistance simultaneously, therefore resulting in significant power enhancement. This result demonstrates that the multilayer structures are useful for improving the thermoelectric performance of the spin Seebeck effect.
\\
\end{abstract}


\maketitle
\begin{figure}[htp!] \center
    \includegraphics[width=17cm, keepaspectratio]{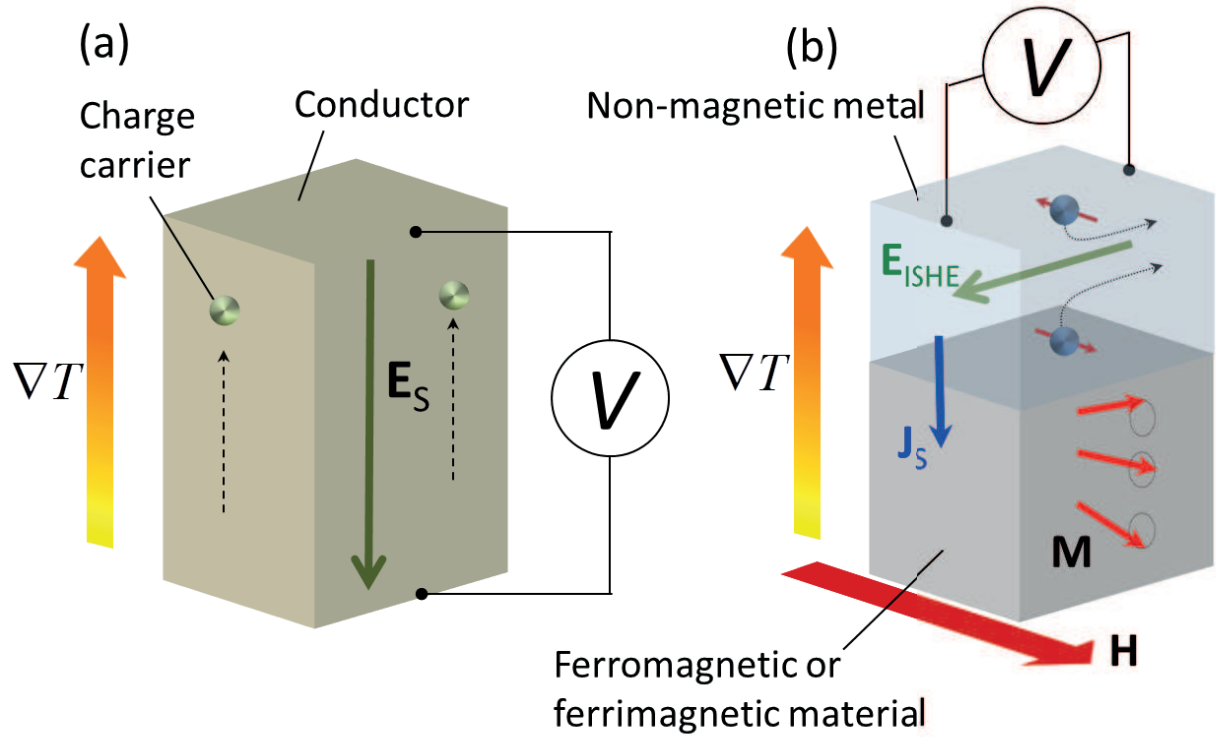}
  \caption{(Color online) Schematic illustrations of (a) the conventional Seebeck effect and (b) the spin Seebeck effect (SSE) with the induced inverse spin Hall effect (ISHE) in a non-magnetic metal attached to a ferromagnetic or ferrimagnetic material. $\mathbf{E}_{\text{S}}$ and $\mathbf{E}_{\text{ISHE}}$ denote the electric fields generated by the Seebeck effect and ISHE, respectively. $\mathbf{H}$, $\mathbf{M}$, $\nabla T$, and $\mathbf{J}_{\text{S}}$ denote the magnetic field, magnetization, temperature gradient, and spatial direction of the spin current, respectively.}
  \label{fig1}
\end{figure}

\begin{figure}[htp!] \center
    \includegraphics[width=17cm, keepaspectratio]{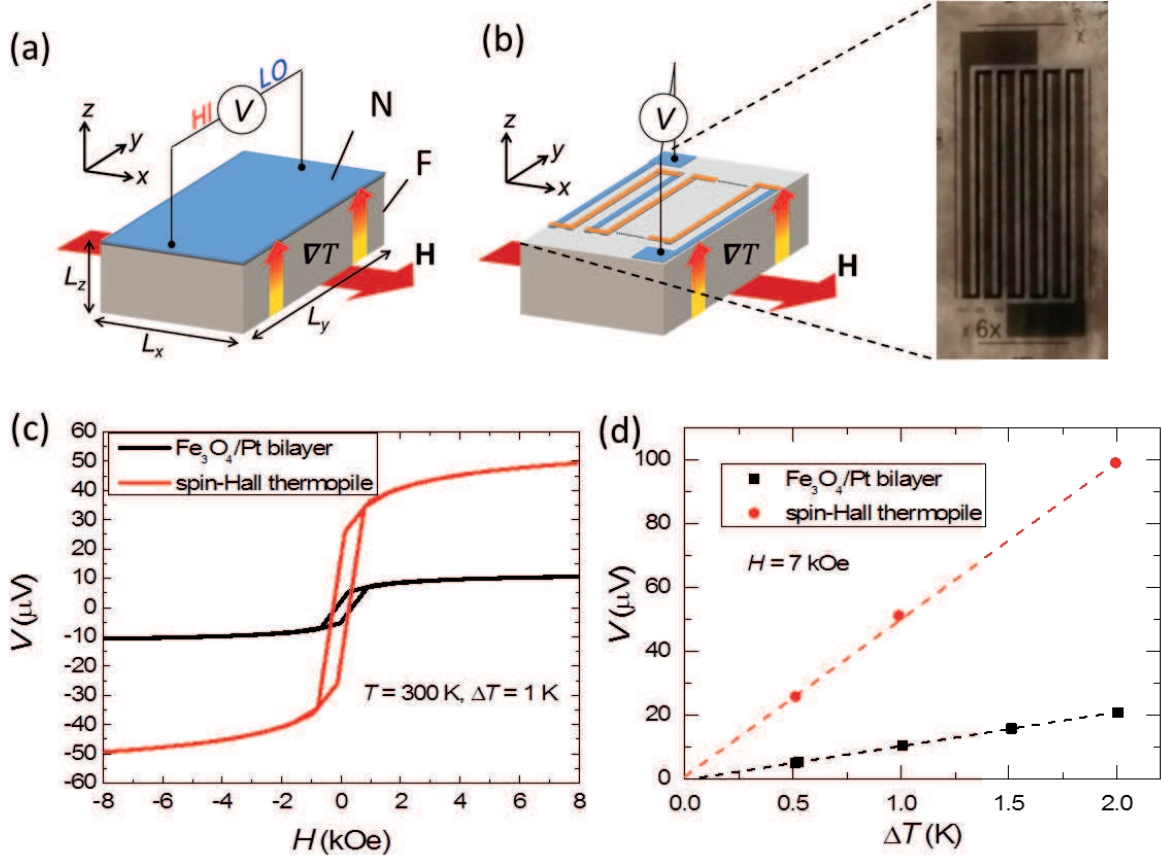}
  \caption{(Color online) Schematic illustrations of the longitudinal SSE measurement configurations (a) in a ferromagnetic insulator(F)/non-magnetic metal(N) bilayer system and (b) in a spin Hall thermopile (an optical image of the thermopile sample is also shown). (c) Magnetic field $H$ dependence of the SSE voltage $V$ response at the temperature difference of $\Delta T$ = 1 K in the simple bilayer and spin Hall thermopile samples comprising a Fe$_3$O$_4$(50)/Pt(5) (thickness in nm) junction. (d) $\Delta T$ dependence of $V$ at $H$ = 7 kOe.}
  \label{fig2}
\end{figure}

\begin{figure}[htp!] \center
    \includegraphics[width=9cm,keepaspectratio]{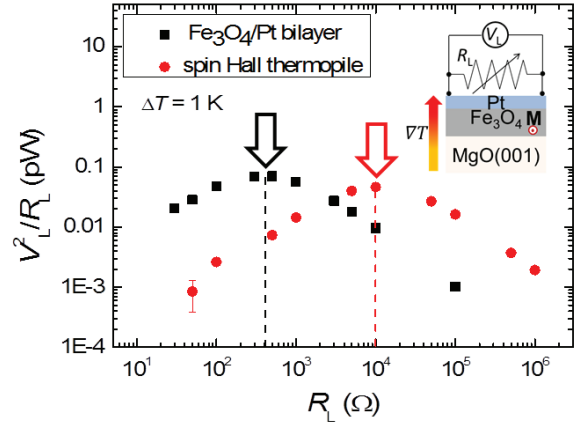}
  \caption{(Color online) Power output ($V^2_\text{L}$/$R_\text{L}$) as a function of the load resistance $R_\text{L}$ obtained for the single bilayer and spin Hall thermopile based on the Fe$_3$O$_4$(50)/Pt(5) junction. $V_\text{L}$ denotes the maximum output power is obtained in the impedance matching condition (see empty arrows). The inset shows the schematic illustration of the measurement configuration to determine the electrical power output characteristics of the SSE.}
  \label{fig3}
\end{figure}

\begin{figure} [htp!]
  \begin{center}
    \includegraphics[width=17cm,keepaspectratio]{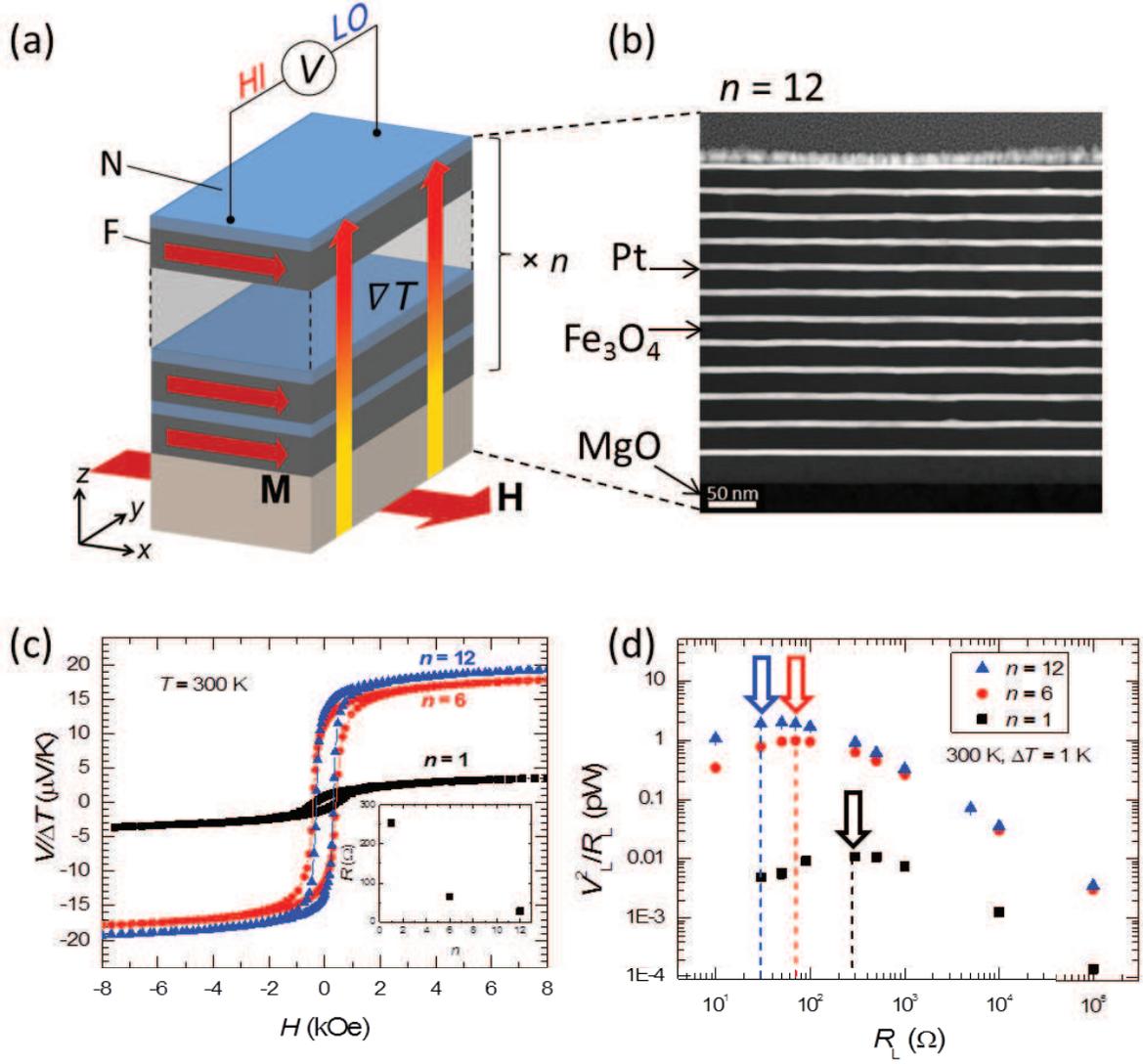}
  \end{center}
  \caption{(Color online) (a) A schematic illustration of the [F/N]$_n$ multilayers, $n$ denotes the number of Fe$_3$O$_4$/Pt bilayers. (b) A scanning transmission electron microscopy image of the cross section of a [Fe$_3$O$_4$(23)/Pt(7)]$_{12}$ sample. (c) $H$ dependence of $V$/$\Delta T$  in the [Fe$_3$O$_4$(23)/Pt(7)]$_n$ multilayers for $n$ = 1, 6 and 12. The inset to (c) shows the resistance of the multilayers as a function of $n$. (d) $V^2_\text{L}$/$R_\text{L}$ in the [Fe$_3$O$_4$(23)/Pt(7)]$_{n}$ multilayers as a function of $R_\text{L}$. Arrows represent the impedance matching condition.} \label{fig4}
\end{figure}

 
Thermoelectrics is one of the promising energy harvesting technologies for waste heat management and recovery, having potential applications in thermoelectric generation or temperature sensing. Thermoelectric generation usually relies on the Seebeck effect, a phenomenon that refers to the generation of an electromotive force parallel to the direction of an applied temperature gradient in electrically conducting materials [see Fig. \ref{fig1}(a)].\\ 
In spintronics,\cite{Wolf2001, Zutic2004a} a different thermoelectric principle was recently discovered. The principle is based on a spin counterpart of the Seebeck effect, in analogy named the spin Seebeck effect (SSE).\cite{uchida:nat2008} The SSE refers to the generation of spin currents \cite{Maekawa2013} in a ferromagnetic material (F) upon application of a temperature gradient. The spin current induced by the SSE is injected into a non-magnetic metal (N) in contact with F and converted into an electromotive force by means of the inverse spin Hall effect (ISHE) in suitable N materials\cite{Saitoh2006a, Valenzuela2006, Kimura2007a, Costache2006} [Fig. \ref{fig1}(b)]. The SSE has been observed in a range of ferromagnetic materials: metals,\cite{uchida:nat2008} semiconductors \cite{jaworsky:natmat2010, jaworsky:prl} and insulators.\cite{uchida:sse-insulator, Meier2013, Niizeki2015} Thus, the SSE has been experimentally established as a general transport phenomenon in ferromagnets. Since the discovery of the SSE, the study of the interaction among spin, charge, and heat currents has been strongly invigorated, which is the main focus of spin caloritronics.\cite{bauer:spinCaloritronics, Boona2014} This has resulted in the discovery of various thermospin effects, such as the spin Peltier effect \cite{Flipse2014} and the spin dependent versions of the ordinary Seebeck \cite{spin-depSeebeck} and Peltier \cite{Flipse2012} effects among others.\cite{LeBreton2011a, Yu2010, An2013c}\\
Theoretically, the SSE \cite{Xiao2010a, Adachi2011, Adachi2013} is understood as an effect resulting from the thermal non-equilibrium between magnons in F and conduction electrons in N, generating a spin current at the F/N interface proportional to the effective temperature difference between magnons in F and electrons in N. Theories from other view points have also been developed,\cite{Hoffman2013, Zhang2012, Rezende2014} where roles of magnon spin currents inside the ferromagnet has been highlighted.\cite{Zhang2012, Rezende2014} This is supported by several recent experiments: the dependence of SSE voltage with the thickness of F,\cite{Kehlberger2015} the enhancement of the SSE voltage in magnetic multilayers,\cite{Ramos2015} and the suppression of the SSE at high magnetic fields.\cite{Kikkawa2015, Jin2015}\\ 
The SSE has potential advantages over conventional thermoelectric technologies for thermal energy harvesting. The experimental geometry, having paths for heat and electric currents independent and perpendicular to each other, allows to have two different materials that can be optimised independently and it is also advantageous for the implementation of thin film devices over large surfaces, for example, simply by thin film coating.\cite{Kirihara2012} Moreover, the observation of the SSE in magnetic insulators implies the potential to generate pure spin currents with lesser dissipation losses due to mobile charge carriers and further expanding the range of possible materials where spin mediated thermoelectric conversion can be studied. However, a main disadvantage of the SSE is the low magnitude of the thermoelectric output. Strategies to overcome this limitation are currently being explored. One possibility is to increase the spin current detection efficiency by taking advantage of the spin Hall angle characteristics of different N materials,\cite{Liu2012, Pai2012, Niimi2012, Laczkowski2014} as it has been shown in the spin Hall thermopiles.\cite{Uchida2012d} A different approach can be directed towards increasing the thermal spin current generation, as recently shown in the SSE of magnetic multilayers.\cite{Ramos2015}\\
In this paper, we have investigated the thermoelectric performance of two types of SSE devices: spin Hall thermopiles \cite{Uchida2012d} and magnetic multilayers formed by repeated growth of $n$ number of [F/N] bilayers, hereafter referred to as [F/N]$_n$. We have performed room temperature measurements of the thermoelectric voltage and power output characteristics for both the structures. The materials chosen to form the basic F/N bilayer structure are magnetite, Fe$_3$O$_4$, as the F layer and platinum, Pt, as the N layer. Magnetite is a ferrimagnetic oxide, commonly employed in spintronics\cite{Wolf2001, Zutic2004a} due to its half-metallic character and high Curie temperature.\cite{Dedkov2002, Walz2002a} These properties have inspired several studies of its magnetotransport properties.\cite{Li1998, Ramos2008, Ramos2014a, Wu2010, Wu2012a} Platinum is a material commonly used for spin current detection, due to its relatively large spin Hall angle.\cite{Hoffmann2013, Sinova2015}\\
The Fe$_3$O$_4$ films were grown on magnesium oxide, MgO(001), substrates by pulsed laser deposition using a KrF excimer laser with 248 nm wavelength, 10 Hz repetition rate, and 3 $\times$ 10$^9$ W/cm$^2$ irradiance in an ultrahigh-vacuum chamber (UHV). The Pt films were deposited in the same UHV chamber by DC magnetron sputtering. Further details on the growth can be found in Ref. 47.\nocite{Orna2010} The film thickness was measured by x-ray reflectivity and its structural quality was confirmed by x-ray diffraction and transmission electron microscopy (TEM). The film cross sections were prepared by Focused Ion Beam and measured by high angle annular dark field (HAADF) scanning transmission electron microscopy (STEM). The measurements were carried out in a probe aberration corrected FEI Titan 60-300 operated at 300 kV. The fabrication of the spin Hall thermopile structure was carried out using  optical lithography, following a two-step process consisting of Ar milling of the Fe$_3$O$_4$/Pt bilayer and fabrication of Al interconnection wires by metal evaporation and lift-off.\\
The SSE experiments were performed using the longitudinal configuration.\cite{Uchida2010i} The sample dimensions are $L_x$ = 2 mm, $L_y$ = 7 mm and $L_z$ = 0.5 mm, except for the spin Hall thermopile structure in which the in plane sample geometry is defined by the lithography pattern. Figure \ref{fig2}(a) shows a schematic illustration of the longitudinal SSE measurement geometry. The sample is placed between two AlN plates with high thermal conductivity, where the top plate has a resistive heater and the lower plate works as a heat sink. By passing an electric current to the heater, a temperature gradient ($\nabla T$) is applied in the $z$ direction. This generates a temperature difference ($\Delta T$) between the bottom and top of the sample, which are stabilized at temperatures $T$ and $T + \Delta T$, respectively. All the measurements were carried out at $T$ = 300 K. The temperature difference is monitored by two thermocouples connected differentially. The SSE voltage in the Pt film is measured along the $y$ direction with a nanovoltmeter, while a magnetic field is applied in the $x$ direction.\\ 
The measurement mechanism is as follows. In the longitudinal configuration, a spin current is injected into the Pt layer with the spatial direction $\mathbf{J}_{\text{S}}$ perpendicular to the Fe$_3$O$_4$/Pt interface (parallel to $\nabla T$) and the spin-polarization vector $\bm{\sigma}$ (parallel to the magnetization of Fe$_3$O$_4$). Then, an electric field ($\mathbf{E}_{\text{ISHE}}$) is generated in Pt due to the ISHE, which can be expressed as:
\begin{equation}
\label{eq:Eishe}
  \mathbf{{E}}_{\text{ISHE}}=\theta_{\text{SH}}\rho(\mathbf{J}_{\text{S}} \times \bm{\sigma}),
\end{equation}
where $\theta_{\text{SH}}$ and $\rho$ denote the spin Hall angle and electric resistivity of Pt, respectively. By measuring the voltage $V$ between the ends of the Pt film, we can electrically detect the SSE by means of the ISHE as $V_{\text{ISHE}} = E_{\text{ISHE}}L_y$. Although the longitudinal configuration is widely used to measure the SSE in ferrimagnetic insulators, in samples with electrically conductive F layers there can be an additional contribution of the anomalous Nernst effect (ANE) in itinerant ferromagnets.\cite{Ramos2014a} Since Fe$_3$O$_4$ is electrically conductive at room temperature, we need to evaluate the contribution from the ANE of the Fe$_3$O$_4$ film to the measured SSE voltage in the Fe$_3$O$_4$/Pt system. We have previously shown that by considering simple equivalent circuit model,\cite{Ramos2013} the following expression can be obtained: $E_y=\frac{r}{1+r}E_{\text{ANE}}$, which gives the electric field in the Fe$_3$O$_4$/Pt bilayer as a function of the ANE-induced electric field ($E_{\text{ANE}}$) and the shunting parameter $r=\frac{\rho_{\text{Pt}}}{\rho_{\text{Fe$_3$O$_4$}}}\frac{t_{\text{Fe$_3$O$_4$}}}{t_{\text{Pt}}}$, with $\rho_{\text{Pt}}$ ($\rho_{\text{Fe$_3$O$_4$}}$) and $t_{\text{Pt}}$ ($t_{\text{Fe$_3$O$_4$}}$) being the resistivity and thickness of the Pt (Fe$_3$O$_4$) layer, respectively.  In the case of the Fe$_3$O$_4$/Pt system, the ANE of the Fe$_3$O$_4$ film is strongly suppressed due to the fact that the resistivity of Fe$_3$O$_4$ is two orders of magnitude greater than that of Pt. For instance, in the Fe$_3$O$_4$/Pt bilayer with $t_{\text{Fe$_3$O$_4$}}$ = 50 nm and $t_{\text{Pt}}$ = 5 nm, used for the spin Hall thermopile, we obtained that the ANE signal in the Fe$_3$O$_4$/Pt bilayer is reduced to 10 \% of the ANE in a Fe$_3$O$_4$ single film. This gives a contribution from the ANE of the Fe$_3$O$_4$ film to the SSE measured in the Fe$_3$O$_4$/Pt bilayer system, which is of only about  2 \% of the measured voltage. Therefore, the SSE dominates the thermoelectric voltage in our Fe$_3$O$_4$/Pt systems.\\
First, we show the SSE in the spin Hall thermopile structure in comparison to that in a single Fe$_3$O$_4$/Pt bilayer. In analogy to thermopile structures commonly used in conventional thermoelectric modules, where the output voltage is increased by connecting a number of thermocouple junctions of materials with dissimilar Seebeck coefficients electrically in series and thermally in parallel,\cite{Snyder2008a}  the spin Hall thermopiles improve the electrical detection of the SSE by enhancing the ISHE voltage. This is achieved by using two materials with different spin Hall angles. These are placed alternatively on top of the ferromagnet forming an array and connected in series forming a zig-zag structure [see Fig. \ref{fig2}(b)].\\
In this study, we fabricated the spin Hall thermopile comprising a set of 6 parallel Fe$_3$O$_4$/Pt wires. The wires were formed by patterning the Fe$_3$O$_4$/Pt bilayer film. This is due to the fact that the ANE suppression depends on the relative resistance values between the Fe$_3$O$_4$ and Pt layers, therefore we cannot deposit an array of Pt wires on a plain Fe$_3$O$_4$ film, as the Pt wires will become comparatively more resistive and the ANE contribution, with respect to  the Fe$_3$O$_4$/Pt bilayer, will change.  The Fe$_3$O$_4$/Pt wires are then electrically connected in series forming a zig-zag structure using a metal with negligible spin Hall angle \cite{Hoffmann2013} and therefore no ISHE contribution (130 nm of Al in our case). In this spin Hall thermopile, the SSE signal is generated by the Fe$_3$O$_4$/Pt wires and the Al wires are only used for electrical connection. The width of the Fe$_3$O$_4$/Pt and Al wires are 200 $\mu$m and 100 $\mu$m, respectively and their length is 5 mm except for the rightmost and leftmost Fe$_3$O$_4$/Pt wires, which have a length of 6 mm including electrode pads [see Fig. \ref{fig2}(b) for an optical image of the thermopile].\\
Figure \ref{fig2}(c) shows the comparative results for the field dependence of the transversal voltage $V$ for the single Fe$_3$O$_4$/Pt bilayer sample and the Fe$_3$O$_4$/Pt spin Hall thermopile at $\Delta T$ = 1 K. We clearly observe a strong enhancement of the SSE voltage in the spin Hall thermopile, as expected from the series connection of the Fe$_3$O$_4$/Pt elements, effectively increasing the length of the sample.  Figure \ref{fig2}(d) shows $V$ at $H$ = 7 kOe as a function of $\Delta T$, thus confirming the linear dependence between $V$ and $\Delta T$. From the slope of these curves we can extract the magnitude of the SSE thermopower of the single bilayer sample and spin Hall thermopile, obtaining the values of 10 $\mu$V/K and 50 $\mu$V/K, respectively. The SSE voltage per one Fe$_3$O$_4$/Pt wire in the spin Hall thermopile is estimated to be 8.2 $\mu$V/K and a value of the electric field per unit temperature difference ($V$/$\Delta TL$) of 1.6 mV/Km is obtained. This value is consistent with the electric field ($V$/$\Delta TL_y$) measured in the Fe$_3$O$_4$/Pt bilayer, therefore, showing that the increase in voltage is proportional to the increase in the effective length of the sample. The thermoelectric voltage of the spin Hall thermopile can be further enhanced by increasing the integration density of wires.\cite{Uchida2012d}\\
The electric power output characteristics of the SSE in the Fe$_3$O$_4$/Pt single bilayer and spin Hall thermopile were also investigated. This is performed by attaching a variable load resistance ($R_\text{L}$) to the samples and measuring the voltage drop across $R_\text{L}$, as schematically shown in the inset of Fig. \ref{fig3}. Then, we estimate the output power as $V^2_\text{L}$/$R_\text{L}$. Figure \ref{fig3} shows the comparative results obtained for the single bilayer and spin Hall thermopile as a function of $R_\text{L}$; the output power is maximized when $R_\text{L}$ equals the internal resistance of the sample, as expected from impedance-matching criterion for maximum power transfer and also in agreement with previous reports.\cite{Uchida2014a} We can see that, despite the increased voltage response in the spin Hall thermopile structure, there is no advantage in terms of extractable electrical power output from the device when compared to the response of the single Fe$_3$O$_4$/Pt bilayer. This is a consequence of the fact that the voltage enhancement is due to the increase in effective length of the sample, which is associated to an increase in the internal resistance.\cite{Uchida2014a}\\
In the case of ANE-based thermopile structures \cite{Sakuraba2013} it was proposed that, if the ANE voltage is independent of sample thickness, the performance can be improved by reducing the internal resistance of the devices, by just increasing the ferromagnet thickness. However, this approach cannot be used in the case of spin Hall thermopiles, since increasing the thickness of the N layers will result in a reduction of the ISHE voltage \cite{Kikkawa_PRB_2013}. Therefore, in order to increase the electric output power of the SSE, we must follow a different approach. To realize the enhancement of the output power of the SSE, we have investigated the thermoelectric performance of the SSE multilayers ([F/N]$_n$) formed by repeated growth of $n$ number of F/N bilayers [Fig. \ref{fig4}(a)]. This type of structure has recently been reported to present an enhancement of the transversal thermoelectric voltage in various magnetic multilayer systems based on oxide/metal,\cite{Ramos2015} all-metallic,\cite{Lee2015, Uchida_2015} and all-oxide \cite{Shiomi2015} junctions. The advantage of the multilayer structure is that along with the voltage enhancement, the internal resistance of the structure decreases with increasing the layer number, due to parallel connection between the layers. Therefore, a significant enhancement of the extractable electric power is expected.\\ 
Here, we show the measured results of voltage and output power characteristics of the SSE in [Fe$_3$O$_4$(23)/Pt(7)]$_n$ multilayer structures, with $n$ = 1, 6 and 12. The structural quality of the multilayers was confirmed by TEM. A STEM-HAADF image of the [Fe$_3$O$_4$/Pt]$_{12}$ multilayer shows well defined interfaces [see Fig. \ref{fig4}(b)]. As shown in Fig. \ref{fig4}(c), the voltage in the multilayer structure with $n$ = 6 and 12 presents an enhancement of about one order of magnitude in comparison with that in the single bilayer with $n$ = 1. The observed enhancement of the SSE voltage in the Fe$_3$O$_4$/Pt multilayer systems can be explained as a result of an increase in the thermally generated spin current flowing across the multilayers.\cite{Ramos2015} This enhancement can be measured due to the presence of Pt interlayers that detect the spin current enhancement within the multilayer. Importantly, since the resistance of the multilayer decreases with increasing the number of the Fe$_3$O$_4$/Pt layers [see inset of Fig \ref{fig4}(c)], it is expected that the power as well as voltage is enhanced. The power output characteristics of the multilayer samples are shown in Fig. \ref{fig4}(d). We found that the power largely increases in the multilayer systems, reaching more than two orders of magnitude enhancement compared to the single bilayer case: the power is estimated to be 0.01, 1 and 2 pW/K$^2$ for $n$ = 1, 6 and 12, respectively.\\
In summary, we have performed thermoelectric voltage and power measurements of the spin Hall thermopile and SSE multilayer structures based on Fe$_3$O$_4$/Pt junction systems. Our results show that the spin Hall thermopile can enhance the voltage, but cannot improve the power output due to the increase of the internal resistance of the devices. However, the voltage increase is beneficial to increase the sensitivity of the SSE and might have possible sensor applications. In contrast, the multilayer SSE devices significantly enhance both thermoelectric voltage and power compared to the conventional bilayer systems, leading to an efficiency improvement in SSE-based thermoelectric generation. Although the obtained power is still small, we believe that the SSE in multilayers can be further enhanced by careful choice of materials and device geometry. Since the observed SSE enhancement in multilayers is expected to depend on spin transport properties across the multilayers, using materials with longer spin transport characteristic lengths might lead to further enhancement in the SSE.\\ 
The authors thank S. Maekawa, H. Adachi, T. Kikkawa, J. Shiomi, T. Oyake, A. Kirihara, and M. Ishida for valuable discussions. The microscopy works were conducted in the Laboratorio de Microscop\'{i}as Avanzadas at Instituto de Nanociencia de Arag\'{o}n, Universidad de Zaragoza. This work was supported by the Spanish Ministry of Science (through projects PRI-PIBJP-2011-0794, MAT2011-27553-C02, including FEDER funding), the Arag\'{o}n Regional Government (project E26), Thermo-spintronic Marie-Curie CIG (Grant Agreement No. 304043), JST-PRESTO ``Phase Interfaces for Highly Efficient Energy Utilization'' from JST, Japan, Grant-in-Aid for Scientific Research on Innovative Areas ``Nano Spin Conversion Science'' (26103005), Grant-in-Aid for Challenging Exploratory Research (26600067), Grant-in-Aid for Scientific Research (A) (15H02012) from MEXT, Japan, NEC Corporation, and The Noguchi Institute.

\bibliographystyle{apsrev}

\begin{thebibliography}{56}
\expandafter\ifx\csname natexlab\endcsname\relax\def\natexlab#1{#1}\fi
\expandafter\ifx\csname bibnamefont\endcsname\relax
  \def\bibnamefont#1{#1}\fi
\expandafter\ifx\csname bibfnamefont\endcsname\relax
  \def\bibfnamefont#1{#1}\fi
\expandafter\ifx\csname citenamefont\endcsname\relax
  \def\citenamefont#1{#1}\fi
\expandafter\ifx\csname url\endcsname\relax
  \def\url#1{\texttt{#1}}\fi
\expandafter\ifx\csname urlprefix\endcsname\relax\def\urlprefix{URL }\fi
\providecommand{\bibinfo}[2]{#2}
\providecommand{\eprint}[2][]{\url{#2}}

\bibitem[{\citenamefont{Wolf et~al.}(2001)\citenamefont{Wolf, Awschalom,
  Buhrman, Daughton, von Moln{\'{a}}r, Roukes, Chtchelkanova, and
  Treger}}]{Wolf2001}
\bibinfo{author}{\bibfnamefont{S.~A.} \bibnamefont{Wolf}},
  \bibinfo{author}{\bibfnamefont{D.~D.} \bibnamefont{Awschalom}},
  \bibinfo{author}{\bibfnamefont{R.~A.} \bibnamefont{Buhrman}},
  \bibinfo{author}{\bibfnamefont{J.~M.} \bibnamefont{Daughton}},
  \bibinfo{author}{\bibfnamefont{S.}~\bibnamefont{von Moln{\'{a}}r}},
  \bibinfo{author}{\bibfnamefont{M.~L.} \bibnamefont{Roukes}},
  \bibinfo{author}{\bibfnamefont{A.~Y.} \bibnamefont{Chtchelkanova}},
  \bibnamefont{and} \bibinfo{author}{\bibfnamefont{D.~M.}
  \bibnamefont{Treger}}, \bibinfo{journal}{Science}
  \textbf{\bibinfo{volume}{294}}, \bibinfo{pages}{1488} (\bibinfo{year}{2001}).

\bibitem[{\citenamefont{{\v{Z}}uti{\'{c}}
  et~al.}(2004)\citenamefont{{\v{Z}}uti{\'{c}}, Fabian, and {Das
  Sarma}}}]{Zutic2004a}
\bibinfo{author}{\bibfnamefont{I.}~\bibnamefont{{\v{Z}}uti{\'{c}}}},
  \bibinfo{author}{\bibfnamefont{J.}~\bibnamefont{Fabian}}, \bibnamefont{and}
  \bibinfo{author}{\bibfnamefont{S.}~\bibnamefont{{Das Sarma}}},
  \bibinfo{journal}{Rev. Mod. Phys.} \textbf{\bibinfo{volume}{76}},
  \bibinfo{pages}{323} (\bibinfo{year}{2004}).

\bibitem[{\citenamefont{Uchida et~al.}(2008)\citenamefont{Uchida, Takahashi,
  Harii, Ieda, Koshibae, Ando, Maekawa, and Saitoh}}]{uchida:nat2008}
\bibinfo{author}{\bibfnamefont{K.}~\bibnamefont{Uchida}},
  \bibinfo{author}{\bibfnamefont{S.}~\bibnamefont{Takahashi}},
  \bibinfo{author}{\bibfnamefont{K.}~\bibnamefont{Harii}},
  \bibinfo{author}{\bibfnamefont{J.}~\bibnamefont{Ieda}},
  \bibinfo{author}{\bibfnamefont{W.}~\bibnamefont{Koshibae}},
  \bibinfo{author}{\bibfnamefont{K.}~\bibnamefont{Ando}},
  \bibinfo{author}{\bibfnamefont{S.}~\bibnamefont{Maekawa}}, \bibnamefont{and}
  \bibinfo{author}{\bibfnamefont{E.}~\bibnamefont{Saitoh}},
  \bibinfo{journal}{Nature} \textbf{\bibinfo{volume}{455}},
  \bibinfo{pages}{778} (\bibinfo{year}{2008}).

\bibitem[{\citenamefont{Maekawa et~al.}(2013)\citenamefont{Maekawa, Adachi,
  Uchida, Ieda, and Saitoh}}]{Maekawa2013}
\bibinfo{author}{\bibfnamefont{S.}~\bibnamefont{Maekawa}},
  \bibinfo{author}{\bibfnamefont{H.}~\bibnamefont{Adachi}},
  \bibinfo{author}{\bibfnamefont{K.}~\bibnamefont{Uchida}},
  \bibinfo{author}{\bibfnamefont{J.}~\bibnamefont{Ieda}}, \bibnamefont{and}
  \bibinfo{author}{\bibfnamefont{E.}~\bibnamefont{Saitoh}},
  \bibinfo{journal}{J. Phys. Soc. Jpn.} \textbf{\bibinfo{volume}{82}},
  \bibinfo{pages}{102002} (\bibinfo{year}{2013}).

\bibitem[{\citenamefont{Saitoh et~al.}(2006)\citenamefont{Saitoh, Ueda,
  Miyajima, and Tatara}}]{Saitoh2006a}
\bibinfo{author}{\bibfnamefont{E.}~\bibnamefont{Saitoh}},
  \bibinfo{author}{\bibfnamefont{M.}~\bibnamefont{Ueda}},
  \bibinfo{author}{\bibfnamefont{H.}~\bibnamefont{Miyajima}}, \bibnamefont{and}
  \bibinfo{author}{\bibfnamefont{G.}~\bibnamefont{Tatara}},
  \bibinfo{journal}{Appl. Phys. Lett.} \textbf{\bibinfo{volume}{88}},
  \bibinfo{pages}{182509} (\bibinfo{year}{2006}).

\bibitem[{\citenamefont{Valenzuela and Tinkham}(2006)}]{Valenzuela2006}
\bibinfo{author}{\bibfnamefont{S.~O.} \bibnamefont{Valenzuela}}
  \bibnamefont{and} \bibinfo{author}{\bibfnamefont{M.}~\bibnamefont{Tinkham}},
  \bibinfo{journal}{Nature} \textbf{\bibinfo{volume}{442}},
  \bibinfo{pages}{176} (\bibinfo{year}{2006}).

\bibitem[{\citenamefont{Kimura et~al.}(2007)\citenamefont{Kimura, Otani, Sato,
  Takahashi, and Maekawa}}]{Kimura2007a}
\bibinfo{author}{\bibfnamefont{T.}~\bibnamefont{Kimura}},
  \bibinfo{author}{\bibfnamefont{Y.}~\bibnamefont{Otani}},
  \bibinfo{author}{\bibfnamefont{T.}~\bibnamefont{Sato}},
  \bibinfo{author}{\bibfnamefont{S.}~\bibnamefont{Takahashi}},
  \bibnamefont{and} \bibinfo{author}{\bibfnamefont{S.}~\bibnamefont{Maekawa}},
  \bibinfo{journal}{Phys. Rev. Lett.} \textbf{\bibinfo{volume}{98}},
  \bibinfo{pages}{156601} (\bibinfo{year}{2007}).

\bibitem[{\citenamefont{Costache et~al.}(2006)\citenamefont{Costache, Sladkov,
  Watts, {van Der Wal}, and {van Wees}}}]{Costache2006}
\bibinfo{author}{\bibfnamefont{M.~V.} \bibnamefont{Costache}},
  \bibinfo{author}{\bibfnamefont{M.}~\bibnamefont{Sladkov}},
  \bibinfo{author}{\bibfnamefont{S.~M.} \bibnamefont{Watts}},
  \bibinfo{author}{\bibfnamefont{C.~H.} \bibnamefont{{van Der Wal}}},
  \bibnamefont{and} \bibinfo{author}{\bibfnamefont{B.~J.} \bibnamefont{{van
  Wees}}}, \bibinfo{journal}{Phys. Rev. Lett.} \textbf{\bibinfo{volume}{97}},
  \bibinfo{pages}{216603} (\bibinfo{year}{2006}).

\bibitem[{\citenamefont{Jaworski et~al.}(2010)\citenamefont{Jaworski, Yang,
  Mack, Awschalom, Heremans, and Myers}}]{jaworsky:natmat2010}
\bibinfo{author}{\bibfnamefont{C.~M.} \bibnamefont{Jaworski}},
  \bibinfo{author}{\bibfnamefont{J.}~\bibnamefont{Yang}},
  \bibinfo{author}{\bibfnamefont{S.}~\bibnamefont{Mack}},
  \bibinfo{author}{\bibfnamefont{D.~D.} \bibnamefont{Awschalom}},
  \bibinfo{author}{\bibfnamefont{J.~P.} \bibnamefont{Heremans}},
  \bibnamefont{and} \bibinfo{author}{\bibfnamefont{R.~C.} \bibnamefont{Myers}},
  \bibinfo{journal}{Nat Mater} \textbf{\bibinfo{volume}{9}},
  \bibinfo{pages}{898} (\bibinfo{year}{2010}).

\bibitem[{\citenamefont{Jaworski et~al.}(2011)\citenamefont{Jaworski, Yang,
  Mack, Awschalom, Myers, and Heremans}}]{jaworsky:prl}
\bibinfo{author}{\bibfnamefont{C.~M.} \bibnamefont{Jaworski}},
  \bibinfo{author}{\bibfnamefont{J.}~\bibnamefont{Yang}},
  \bibinfo{author}{\bibfnamefont{S.}~\bibnamefont{Mack}},
  \bibinfo{author}{\bibfnamefont{D.~D.} \bibnamefont{Awschalom}},
  \bibinfo{author}{\bibfnamefont{R.~C.} \bibnamefont{Myers}}, \bibnamefont{and}
  \bibinfo{author}{\bibfnamefont{J.~P.} \bibnamefont{Heremans}},
  \bibinfo{journal}{Phys. Rev. Lett.} \textbf{\bibinfo{volume}{106}},
  \bibinfo{pages}{186601} (\bibinfo{year}{2011}).

\bibitem[{\citenamefont{Uchida et~al.}(2010{\natexlab{a}})\citenamefont{Uchida,
  Xiao, Adachi, Ohe, Takahashi, Ieda, Ota, Kajiwara, Umezawa, Kawai
  et~al.}}]{uchida:sse-insulator}
\bibinfo{author}{\bibfnamefont{K.}~\bibnamefont{Uchida}},
  \bibinfo{author}{\bibfnamefont{J.}~\bibnamefont{Xiao}},
  \bibinfo{author}{\bibfnamefont{H.}~\bibnamefont{Adachi}},
  \bibinfo{author}{\bibfnamefont{J.}~\bibnamefont{Ohe}},
  \bibinfo{author}{\bibfnamefont{S.}~\bibnamefont{Takahashi}},
  \bibinfo{author}{\bibfnamefont{J.}~\bibnamefont{Ieda}},
  \bibinfo{author}{\bibfnamefont{T.}~\bibnamefont{Ota}},
  \bibinfo{author}{\bibfnamefont{Y.}~\bibnamefont{Kajiwara}},
  \bibinfo{author}{\bibfnamefont{H.}~\bibnamefont{Umezawa}},
  \bibinfo{author}{\bibfnamefont{H.}~\bibnamefont{Kawai}},
  \bibnamefont{et~al.}, \bibinfo{journal}{Nat. Mater.}
  \textbf{\bibinfo{volume}{9}}, \bibinfo{pages}{894}
  (\bibinfo{year}{2010}{\natexlab{a}}).

\bibitem[{\citenamefont{Meier et~al.}(2013)\citenamefont{Meier, Kuschel, Shen,
  Gupta, Kikkawa, Uchida, Saitoh, Schmalhorst, and Reiss}}]{Meier2013}
\bibinfo{author}{\bibfnamefont{D.}~\bibnamefont{Meier}},
  \bibinfo{author}{\bibfnamefont{T.}~\bibnamefont{Kuschel}},
  \bibinfo{author}{\bibfnamefont{L.}~\bibnamefont{Shen}},
  \bibinfo{author}{\bibfnamefont{A.}~\bibnamefont{Gupta}},
  \bibinfo{author}{\bibfnamefont{T.}~\bibnamefont{Kikkawa}},
  \bibinfo{author}{\bibfnamefont{K.}~\bibnamefont{Uchida}},
  \bibinfo{author}{\bibfnamefont{E.}~\bibnamefont{Saitoh}},
  \bibinfo{author}{\bibfnamefont{J.-M.} \bibnamefont{Schmalhorst}},
  \bibnamefont{and} \bibinfo{author}{\bibfnamefont{G.}~\bibnamefont{Reiss}},
  \bibinfo{journal}{Physical Review B} \textbf{\bibinfo{volume}{87}},
  \bibinfo{pages}{054421} (\bibinfo{year}{2013}).

\bibitem[{\citenamefont{Niizeki et~al.}(2015)\citenamefont{Niizeki, Kikkawa,
  Uchida, Oka, Suzuki, Yanagihara, Kita, and Saitoh}}]{Niizeki2015}
\bibinfo{author}{\bibfnamefont{T.}~\bibnamefont{Niizeki}},
  \bibinfo{author}{\bibfnamefont{T.}~\bibnamefont{Kikkawa}},
  \bibinfo{author}{\bibfnamefont{K.}~\bibnamefont{Uchida}},
  \bibinfo{author}{\bibfnamefont{M.}~\bibnamefont{Oka}},
  \bibinfo{author}{\bibfnamefont{K.~Z.} \bibnamefont{Suzuki}},
  \bibinfo{author}{\bibfnamefont{H.}~\bibnamefont{Yanagihara}},
  \bibinfo{author}{\bibfnamefont{E.}~\bibnamefont{Kita}}, \bibnamefont{and}
  \bibinfo{author}{\bibfnamefont{E.}~\bibnamefont{Saitoh}},
  \bibinfo{journal}{AIP Advances} \textbf{\bibinfo{volume}{5}},
  \bibinfo{pages}{053603} (\bibinfo{year}{2015}).

\bibitem[{\citenamefont{Bauer et~al.}(2012)\citenamefont{Bauer, Saitoh, and van
  Wees}}]{bauer:spinCaloritronics}
\bibinfo{author}{\bibfnamefont{G.~E.~W.} \bibnamefont{Bauer}},
  \bibinfo{author}{\bibfnamefont{E.}~\bibnamefont{Saitoh}}, \bibnamefont{and}
  \bibinfo{author}{\bibfnamefont{B.~J.} \bibnamefont{van Wees}},
  \bibinfo{journal}{Nat. Mater.} \textbf{\bibinfo{volume}{11}},
  \bibinfo{pages}{391} (\bibinfo{year}{2012}).

\bibitem[{\citenamefont{Boona et~al.}(2014)\citenamefont{Boona, Myers, and
  Heremans}}]{Boona2014}
\bibinfo{author}{\bibfnamefont{S.~R.} \bibnamefont{Boona}},
  \bibinfo{author}{\bibfnamefont{R.~C.} \bibnamefont{Myers}}, \bibnamefont{and}
  \bibinfo{author}{\bibfnamefont{J.~P.} \bibnamefont{Heremans}},
  \bibinfo{journal}{Energy Environ. Sci.} \textbf{\bibinfo{volume}{7}},
  \bibinfo{pages}{885} (\bibinfo{year}{2014}).

\bibitem[{\citenamefont{Flipse et~al.}(2014)\citenamefont{Flipse, Dejene,
  Wagenaar, Bauer, Youssef, and Wees}}]{Flipse2014}
\bibinfo{author}{\bibfnamefont{J.}~\bibnamefont{Flipse}},
  \bibinfo{author}{\bibfnamefont{F.~K.} \bibnamefont{Dejene}},
  \bibinfo{author}{\bibfnamefont{D.}~\bibnamefont{Wagenaar}},
  \bibinfo{author}{\bibfnamefont{G.~E.~W.} \bibnamefont{Bauer}},
  \bibinfo{author}{\bibfnamefont{J.~B.} \bibnamefont{Youssef}},
  \bibnamefont{and} \bibinfo{author}{\bibfnamefont{B.~J.~V.}
  \bibnamefont{Wees}}, \bibinfo{journal}{Phys. Rev. Lett.}
  \textbf{\bibinfo{volume}{113}}, \bibinfo{pages}{027601}
  (\bibinfo{year}{2014}).

\bibitem[{\citenamefont{Slachter et~al.}(2010)\citenamefont{Slachter, Bakker,
  Adam, and van Wees}}]{spin-depSeebeck}
\bibinfo{author}{\bibfnamefont{A.}~\bibnamefont{Slachter}},
  \bibinfo{author}{\bibfnamefont{F.~L.} \bibnamefont{Bakker}},
  \bibinfo{author}{\bibfnamefont{J.-P.} \bibnamefont{Adam}}, \bibnamefont{and}
  \bibinfo{author}{\bibfnamefont{B.~J.} \bibnamefont{van Wees}},
  \bibinfo{journal}{Nat. Phys.} \textbf{\bibinfo{volume}{6}},
  \bibinfo{pages}{879} (\bibinfo{year}{2010}).

\bibitem[{\citenamefont{Flipse et~al.}(2012)\citenamefont{Flipse, Bakker,
  Slachter, Dejene, and van Wees}}]{Flipse2012}
\bibinfo{author}{\bibfnamefont{J.}~\bibnamefont{Flipse}},
  \bibinfo{author}{\bibfnamefont{F.~L.} \bibnamefont{Bakker}},
  \bibinfo{author}{\bibfnamefont{a.}~\bibnamefont{Slachter}},
  \bibinfo{author}{\bibfnamefont{F.~K.} \bibnamefont{Dejene}},
  \bibnamefont{and} \bibinfo{author}{\bibfnamefont{B.~J.} \bibnamefont{van
  Wees}}, \bibinfo{journal}{Nat. Nanotechnol.} \textbf{\bibinfo{volume}{7}},
  \bibinfo{pages}{166} (\bibinfo{year}{2012}).

\bibitem[{\citenamefont{{Le Breton} et~al.}(2011)\citenamefont{{Le Breton},
  Sharma, Saito, Yuasa, and Jansen}}]{LeBreton2011a}
\bibinfo{author}{\bibfnamefont{J.-C.} \bibnamefont{{Le Breton}}},
  \bibinfo{author}{\bibfnamefont{S.}~\bibnamefont{Sharma}},
  \bibinfo{author}{\bibfnamefont{H.}~\bibnamefont{Saito}},
  \bibinfo{author}{\bibfnamefont{S.}~\bibnamefont{Yuasa}}, \bibnamefont{and}
  \bibinfo{author}{\bibfnamefont{R.}~\bibnamefont{Jansen}},
  \bibinfo{journal}{Nature} \textbf{\bibinfo{volume}{475}}, \bibinfo{pages}{82}
  (\bibinfo{year}{2011}).

\bibitem[{\citenamefont{Yu et~al.}(2010)\citenamefont{Yu, Granville, Yu, and
  Ansermet}}]{Yu2010}
\bibinfo{author}{\bibfnamefont{H.}~\bibnamefont{Yu}},
  \bibinfo{author}{\bibfnamefont{S.}~\bibnamefont{Granville}},
  \bibinfo{author}{\bibfnamefont{D.~P.} \bibnamefont{Yu}}, \bibnamefont{and}
  \bibinfo{author}{\bibfnamefont{J.-P.} \bibnamefont{Ansermet}},
  \bibinfo{journal}{Phys. Rev. Lett.} \textbf{\bibinfo{volume}{104}},
  \bibinfo{pages}{146601} (\bibinfo{year}{2010}).

\bibitem[{\citenamefont{An et~al.}(2013)\citenamefont{An, Vasyuchka, Uchida,
  Chumak, Yamaguchi, Harii, Ohe, Jungfleisch, Kajiwara, Adachi
  et~al.}}]{An2013c}
\bibinfo{author}{\bibfnamefont{T.}~\bibnamefont{An}},
  \bibinfo{author}{\bibfnamefont{V.~I.} \bibnamefont{Vasyuchka}},
  \bibinfo{author}{\bibfnamefont{K.}~\bibnamefont{Uchida}},
  \bibinfo{author}{\bibfnamefont{A.~V.} \bibnamefont{Chumak}},
  \bibinfo{author}{\bibfnamefont{K.}~\bibnamefont{Yamaguchi}},
  \bibinfo{author}{\bibfnamefont{K.}~\bibnamefont{Harii}},
  \bibinfo{author}{\bibfnamefont{J.}~\bibnamefont{Ohe}},
  \bibinfo{author}{\bibfnamefont{M.~B.} \bibnamefont{Jungfleisch}},
  \bibinfo{author}{\bibfnamefont{Y.}~\bibnamefont{Kajiwara}},
  \bibinfo{author}{\bibfnamefont{H.}~\bibnamefont{Adachi}},
  \bibnamefont{et~al.}, \bibinfo{journal}{Nat. Mat.}
  \textbf{\bibinfo{volume}{12}}, \bibinfo{pages}{549} (\bibinfo{year}{2013}).

\bibitem[{\citenamefont{Xiao et~al.}(2010)\citenamefont{Xiao, Bauer, Uchida,
  Saitoh, and Maekawa}}]{Xiao2010a}
\bibinfo{author}{\bibfnamefont{J.}~\bibnamefont{Xiao}},
  \bibinfo{author}{\bibfnamefont{G.~E.~W.} \bibnamefont{Bauer}},
  \bibinfo{author}{\bibfnamefont{K.}~\bibnamefont{Uchida}},
  \bibinfo{author}{\bibfnamefont{E.}~\bibnamefont{Saitoh}}, \bibnamefont{and}
  \bibinfo{author}{\bibfnamefont{S.}~\bibnamefont{Maekawa}},
  \bibinfo{journal}{Phys. Rev. B} \textbf{\bibinfo{volume}{81}},
  \bibinfo{pages}{214418} (\bibinfo{year}{2010}).

\bibitem[{\citenamefont{Adachi et~al.}(2011)\citenamefont{Adachi, Ohe,
  Takahashi, and Maekawa}}]{Adachi2011}
\bibinfo{author}{\bibfnamefont{H.}~\bibnamefont{Adachi}},
  \bibinfo{author}{\bibfnamefont{J.-i.} \bibnamefont{Ohe}},
  \bibinfo{author}{\bibfnamefont{S.}~\bibnamefont{Takahashi}},
  \bibnamefont{and} \bibinfo{author}{\bibfnamefont{S.}~\bibnamefont{Maekawa}},
  \bibinfo{journal}{Phys. Rev. B} \textbf{\bibinfo{volume}{83}},
  \bibinfo{pages}{094410} (\bibinfo{year}{2011}).

\bibitem[{\citenamefont{Adachi et~al.}(2013)\citenamefont{Adachi, Uchida,
  Saitoh, and Maekawa}}]{Adachi2013}
\bibinfo{author}{\bibfnamefont{H.}~\bibnamefont{Adachi}},
  \bibinfo{author}{\bibfnamefont{K.}~\bibnamefont{Uchida}},
  \bibinfo{author}{\bibfnamefont{E.}~\bibnamefont{Saitoh}}, \bibnamefont{and}
  \bibinfo{author}{\bibfnamefont{S.}~\bibnamefont{Maekawa}},
  \bibinfo{journal}{Rep. Prog. Phys.} \textbf{\bibinfo{volume}{76}},
  \bibinfo{pages}{036501} (\bibinfo{year}{2013}).

\bibitem[{\citenamefont{Hoffman et~al.}(2013)\citenamefont{Hoffman, Sato, and
  Tserkovnyak}}]{Hoffman2013}
\bibinfo{author}{\bibfnamefont{S.}~\bibnamefont{Hoffman}},
  \bibinfo{author}{\bibfnamefont{K.}~\bibnamefont{Sato}}, \bibnamefont{and}
  \bibinfo{author}{\bibfnamefont{Y.}~\bibnamefont{Tserkovnyak}},
  \bibinfo{journal}{Phys. Rev. B} \textbf{\bibinfo{volume}{88}},
  \bibinfo{pages}{064408} (\bibinfo{year}{2013}).

\bibitem[{\citenamefont{Zhang and Zhang}(2012)}]{Zhang2012}
\bibinfo{author}{\bibfnamefont{S.~S.-L.} \bibnamefont{Zhang}} \bibnamefont{and}
  \bibinfo{author}{\bibfnamefont{S.}~\bibnamefont{Zhang}},
  \bibinfo{journal}{Phys. Rev. B} \textbf{\bibinfo{volume}{86}},
  \bibinfo{pages}{214424} (\bibinfo{year}{2012}).

\bibitem[{\citenamefont{Rezende et~al.}(2014)\citenamefont{Rezende,
  Rodr{\'{i}}guez-Su{\'{a}}rez, Cunha, Rodrigues, Machado, {Fonseca Guerra},
  {Lopez Ortiz}, and Azevedo}}]{Rezende2014}
\bibinfo{author}{\bibfnamefont{S.~M.} \bibnamefont{Rezende}},
  \bibinfo{author}{\bibfnamefont{R.~L.}
  \bibnamefont{Rodr{\'{i}}guez-Su{\'{a}}rez}},
  \bibinfo{author}{\bibfnamefont{R.~O.} \bibnamefont{Cunha}},
  \bibinfo{author}{\bibfnamefont{A.~R.} \bibnamefont{Rodrigues}},
  \bibinfo{author}{\bibfnamefont{F.~L.~A.} \bibnamefont{Machado}},
  \bibinfo{author}{\bibfnamefont{G.~A.} \bibnamefont{{Fonseca Guerra}}},
  \bibinfo{author}{\bibfnamefont{J.~C.} \bibnamefont{{Lopez Ortiz}}},
  \bibnamefont{and} \bibinfo{author}{\bibfnamefont{A.}~\bibnamefont{Azevedo}},
  \bibinfo{journal}{Phys. Rev. B} \textbf{\bibinfo{volume}{89}},
  \bibinfo{pages}{014416} (\bibinfo{year}{2014}).

\bibitem[{\citenamefont{Kehlberger et~al.}(2015)\citenamefont{Kehlberger,
  Ritzmann, Hinzke, Guo, Cramer, Jakob, Onbasli, Kim, Ross, Jungfleisch
  et~al.}}]{Kehlberger2015}
\bibinfo{author}{\bibfnamefont{A.}~\bibnamefont{Kehlberger}},
  \bibinfo{author}{\bibfnamefont{U.}~\bibnamefont{Ritzmann}},
  \bibinfo{author}{\bibfnamefont{D.}~\bibnamefont{Hinzke}},
  \bibinfo{author}{\bibfnamefont{E.-J.} \bibnamefont{Guo}},
  \bibinfo{author}{\bibfnamefont{J.}~\bibnamefont{Cramer}},
  \bibinfo{author}{\bibfnamefont{G.}~\bibnamefont{Jakob}},
  \bibinfo{author}{\bibfnamefont{M.~C.} \bibnamefont{Onbasli}},
  \bibinfo{author}{\bibfnamefont{D.~H.} \bibnamefont{Kim}},
  \bibinfo{author}{\bibfnamefont{C.~A.} \bibnamefont{Ross}},
  \bibinfo{author}{\bibfnamefont{M.~B.} \bibnamefont{Jungfleisch}},
  \bibnamefont{et~al.}, \bibinfo{journal}{Phys. Rev. Lett.}
  \textbf{\bibinfo{volume}{115}}, \bibinfo{pages}{096602}
  (\bibinfo{year}{2015}).

\bibitem[{\citenamefont{Ramos et~al.}(2015)\citenamefont{Ramos, Kikkawa,
  Aguirre, Lucas, Anad{\'{o}}n, Oyake, Uchida, and Adachi}}]{Ramos2015}
\bibinfo{author}{\bibfnamefont{R.}~\bibnamefont{Ramos}},
  \bibinfo{author}{\bibfnamefont{T.}~\bibnamefont{Kikkawa}},
  \bibinfo{author}{\bibfnamefont{M.~H.} \bibnamefont{Aguirre}},
  \bibinfo{author}{\bibfnamefont{I.}~\bibnamefont{Lucas}},
  \bibinfo{author}{\bibfnamefont{A.}~\bibnamefont{Anad{\'{o}}n}},
  \bibinfo{author}{\bibfnamefont{T.}~\bibnamefont{Oyake}},
  \bibinfo{author}{\bibfnamefont{K.}~\bibnamefont{Uchida}}, \bibnamefont{and}
  \bibinfo{author}{\bibfnamefont{H.}~\bibnamefont{Adachi}},
  \bibinfo{journal}{Phys. Rev. B} \textbf{\bibinfo{volume}{92}},
  \bibinfo{pages}{220407(R)} (\bibinfo{year}{2015}).

\bibitem[{\citenamefont{Kikkawa et~al.}(2015)\citenamefont{Kikkawa, Uchida,
  Daimon, Qiu, Shiomi, and Saitoh}}]{Kikkawa2015}
\bibinfo{author}{\bibfnamefont{T.}~\bibnamefont{Kikkawa}},
  \bibinfo{author}{\bibfnamefont{K.}~\bibnamefont{Uchida}},
  \bibinfo{author}{\bibfnamefont{S.}~\bibnamefont{Daimon}},
  \bibinfo{author}{\bibfnamefont{Z.}~\bibnamefont{Qiu}},
  \bibinfo{author}{\bibfnamefont{Y.}~\bibnamefont{Shiomi}}, \bibnamefont{and}
  \bibinfo{author}{\bibfnamefont{E.}~\bibnamefont{Saitoh}},
  \bibinfo{journal}{Phys. Rev. B} \textbf{\bibinfo{volume}{92}},
  \bibinfo{pages}{064413} (\bibinfo{year}{2015}).

\bibitem[{\citenamefont{Jin et~al.}(2015)\citenamefont{Jin, Boona, Yang, Myers,
  and Heremans}}]{Jin2015}
\bibinfo{author}{\bibfnamefont{H.}~\bibnamefont{Jin}},
  \bibinfo{author}{\bibfnamefont{S.~R.} \bibnamefont{Boona}},
  \bibinfo{author}{\bibfnamefont{Z.}~\bibnamefont{Yang}},
  \bibinfo{author}{\bibfnamefont{R.~C.} \bibnamefont{Myers}}, \bibnamefont{and}
  \bibinfo{author}{\bibfnamefont{J.~P.} \bibnamefont{Heremans}},
  \bibinfo{journal}{Phys. Rev. B} \textbf{\bibinfo{volume}{92}},
  \bibinfo{pages}{054436} (\bibinfo{year}{2015}).

\bibitem[{\citenamefont{Kirihara et~al.}(2012)\citenamefont{Kirihara, Uchida,
  Kajiwara, Ishida, Nakamura, Manako, Saitoh, and Yorozu}}]{Kirihara2012}
\bibinfo{author}{\bibfnamefont{A.}~\bibnamefont{Kirihara}},
  \bibinfo{author}{\bibfnamefont{K.}~\bibnamefont{Uchida}},
  \bibinfo{author}{\bibfnamefont{Y.}~\bibnamefont{Kajiwara}},
  \bibinfo{author}{\bibfnamefont{M.}~\bibnamefont{Ishida}},
  \bibinfo{author}{\bibfnamefont{Y.}~\bibnamefont{Nakamura}},
  \bibinfo{author}{\bibfnamefont{T.}~\bibnamefont{Manako}},
  \bibinfo{author}{\bibfnamefont{E.}~\bibnamefont{Saitoh}}, \bibnamefont{and}
  \bibinfo{author}{\bibfnamefont{S.}~\bibnamefont{Yorozu}},
  \textbf{\bibinfo{volume}{11}}, \bibinfo{pages}{686} (\bibinfo{year}{2012}).

\bibitem[{\citenamefont{Liu et~al.}(2012)\citenamefont{Liu, Pai, Li, Tseng,
  Ralph, and Buhrman}}]{Liu2012}
\bibinfo{author}{\bibfnamefont{L.}~\bibnamefont{Liu}},
  \bibinfo{author}{\bibfnamefont{C.-F.} \bibnamefont{Pai}},
  \bibinfo{author}{\bibfnamefont{Y.}~\bibnamefont{Li}},
  \bibinfo{author}{\bibfnamefont{H.~W.} \bibnamefont{Tseng}},
  \bibinfo{author}{\bibfnamefont{D.~C.} \bibnamefont{Ralph}}, \bibnamefont{and}
  \bibinfo{author}{\bibfnamefont{R.~A.} \bibnamefont{Buhrman}},
  \bibinfo{journal}{Science} \textbf{\bibinfo{volume}{336}},
  \bibinfo{pages}{555} (\bibinfo{year}{2012}).

\bibitem[{\citenamefont{Pai et~al.}(2012)\citenamefont{Pai, Liu, Li, Tseng,
  Ralph, and Buhrman}}]{Pai2012}
\bibinfo{author}{\bibfnamefont{C.-F.} \bibnamefont{Pai}},
  \bibinfo{author}{\bibfnamefont{L.}~\bibnamefont{Liu}},
  \bibinfo{author}{\bibfnamefont{Y.}~\bibnamefont{Li}},
  \bibinfo{author}{\bibfnamefont{H.~W.} \bibnamefont{Tseng}},
  \bibinfo{author}{\bibfnamefont{D.~C.} \bibnamefont{Ralph}}, \bibnamefont{and}
  \bibinfo{author}{\bibfnamefont{R.~A.} \bibnamefont{Buhrman}},
  \bibinfo{journal}{Appl. Phys. Lett.} \textbf{\bibinfo{volume}{101}},
  \bibinfo{pages}{122404} (\bibinfo{year}{2012}).

\bibitem[{\citenamefont{Niimi et~al.}(2012)\citenamefont{Niimi, Kawanishi, Wei,
  Deranlot, Yang, Chshiev, Valet, Fert, and Otani}}]{Niimi2012}
\bibinfo{author}{\bibfnamefont{Y.}~\bibnamefont{Niimi}},
  \bibinfo{author}{\bibfnamefont{Y.}~\bibnamefont{Kawanishi}},
  \bibinfo{author}{\bibfnamefont{D.~H.} \bibnamefont{Wei}},
  \bibinfo{author}{\bibfnamefont{C.}~\bibnamefont{Deranlot}},
  \bibinfo{author}{\bibfnamefont{H.~X.} \bibnamefont{Yang}},
  \bibinfo{author}{\bibfnamefont{M.}~\bibnamefont{Chshiev}},
  \bibinfo{author}{\bibfnamefont{T.}~\bibnamefont{Valet}},
  \bibinfo{author}{\bibfnamefont{A.}~\bibnamefont{Fert}}, \bibnamefont{and}
  \bibinfo{author}{\bibfnamefont{Y.}~\bibnamefont{Otani}},
  \bibinfo{journal}{Phys. Rev. Lett.} \textbf{\bibinfo{volume}{109}},
  \bibinfo{pages}{156602} (\bibinfo{year}{2012}).

\bibitem[{\citenamefont{Laczkowski et~al.}(2014)\citenamefont{Laczkowski,
  Rojas-S{\'{a}}nchez, Savero-Torres, Jaffr{\`{e}}s, Reyren, Deranlot, Notin,
  Beign{\'{e}}, Marty, Attan{\'{e}} et~al.}}]{Laczkowski2014}
\bibinfo{author}{\bibfnamefont{P.}~\bibnamefont{Laczkowski}},
  \bibinfo{author}{\bibfnamefont{J.-C.} \bibnamefont{Rojas-S{\'{a}}nchez}},
  \bibinfo{author}{\bibfnamefont{W.}~\bibnamefont{Savero-Torres}},
  \bibinfo{author}{\bibfnamefont{H.}~\bibnamefont{Jaffr{\`{e}}s}},
  \bibinfo{author}{\bibfnamefont{N.}~\bibnamefont{Reyren}},
  \bibinfo{author}{\bibfnamefont{C.}~\bibnamefont{Deranlot}},
  \bibinfo{author}{\bibfnamefont{L.}~\bibnamefont{Notin}},
  \bibinfo{author}{\bibfnamefont{C.}~\bibnamefont{Beign{\'{e}}}},
  \bibinfo{author}{\bibfnamefont{a.}~\bibnamefont{Marty}},
  \bibinfo{author}{\bibfnamefont{J.-P.} \bibnamefont{Attan{\'{e}}}},
  \bibnamefont{et~al.}, \bibinfo{journal}{Appl. Phys. Lett.}
  \textbf{\bibinfo{volume}{104}}, \bibinfo{pages}{142403}
  (\bibinfo{year}{2014}).

\bibitem[{\citenamefont{Uchida et~al.}(2012)\citenamefont{Uchida, Nonaka,
  Yoshino, Kikkawa, Kikuchi, and Saitoh}}]{Uchida2012d}
\bibinfo{author}{\bibfnamefont{K.}~\bibnamefont{Uchida}},
  \bibinfo{author}{\bibfnamefont{T.}~\bibnamefont{Nonaka}},
  \bibinfo{author}{\bibfnamefont{T.}~\bibnamefont{Yoshino}},
  \bibinfo{author}{\bibfnamefont{T.}~\bibnamefont{Kikkawa}},
  \bibinfo{author}{\bibfnamefont{D.}~\bibnamefont{Kikuchi}}, \bibnamefont{and}
  \bibinfo{author}{\bibfnamefont{E.}~\bibnamefont{Saitoh}},
  \bibinfo{journal}{Appl. Phys. Express} \textbf{\bibinfo{volume}{5}},
  \bibinfo{pages}{093001} (\bibinfo{year}{2012}).

\bibitem[{\citenamefont{Dedkov et~al.}(2002)\citenamefont{Dedkov,
  R{\"{u}}diger, and G{\"{u}}ntherodt}}]{Dedkov2002}
\bibinfo{author}{\bibfnamefont{Y.~S.} \bibnamefont{Dedkov}},
  \bibinfo{author}{\bibfnamefont{U.}~\bibnamefont{R{\"{u}}diger}},
  \bibnamefont{and}
  \bibinfo{author}{\bibfnamefont{G.}~\bibnamefont{G{\"{u}}ntherodt}},
  \bibinfo{journal}{Phys. Rev. B} \textbf{\bibinfo{volume}{65}},
  \bibinfo{pages}{64417} (\bibinfo{year}{2002}).

\bibitem[{\citenamefont{Walz}(2002)}]{Walz2002a}
\bibinfo{author}{\bibfnamefont{F.}~\bibnamefont{Walz}}, \bibinfo{journal}{J.
  Phys.: Condens. Matter} \textbf{\bibinfo{volume}{14}}, \bibinfo{pages}{R285}
  (\bibinfo{year}{2002}).

\bibitem[{\citenamefont{Li et~al.}(1998)\citenamefont{Li, Gupta, Xiao, Qian,
  and Dravid}}]{Li1998}
\bibinfo{author}{\bibfnamefont{X.~W.} \bibnamefont{Li}},
  \bibinfo{author}{\bibfnamefont{a.}~\bibnamefont{Gupta}},
  \bibinfo{author}{\bibfnamefont{G.}~\bibnamefont{Xiao}},
  \bibinfo{author}{\bibfnamefont{W.}~\bibnamefont{Qian}}, \bibnamefont{and}
  \bibinfo{author}{\bibfnamefont{V.~P.} \bibnamefont{Dravid}},
  \bibinfo{journal}{Appl. Phys. Lett.} \textbf{\bibinfo{volume}{73}},
  \bibinfo{pages}{3282} (\bibinfo{year}{1998}).

\bibitem[{\citenamefont{Ramos et~al.}(2008)\citenamefont{Ramos, Arora, and
  Shvets}}]{Ramos2008}
\bibinfo{author}{\bibfnamefont{R.}~\bibnamefont{Ramos}},
  \bibinfo{author}{\bibfnamefont{S.~K.} \bibnamefont{Arora}}, \bibnamefont{and}
  \bibinfo{author}{\bibfnamefont{I.~V.} \bibnamefont{Shvets}},
  \bibinfo{journal}{Phys. Rev. B} \textbf{\bibinfo{volume}{78}},
  \bibinfo{pages}{214402} (\bibinfo{year}{2008}).

\bibitem[{\citenamefont{Ramos et~al.}(2014)\citenamefont{Ramos, Aguirre,
  Anad{\'{o}}n, Blasco, Lucas, Uchida, Algarabel, Morell{\'{o}}n, Saitoh, and
  Ibarra}}]{Ramos2014a}
\bibinfo{author}{\bibfnamefont{R.}~\bibnamefont{Ramos}},
  \bibinfo{author}{\bibfnamefont{M.~H.} \bibnamefont{Aguirre}},
  \bibinfo{author}{\bibfnamefont{A.}~\bibnamefont{Anad{\'{o}}n}},
  \bibinfo{author}{\bibfnamefont{J.}~\bibnamefont{Blasco}},
  \bibinfo{author}{\bibfnamefont{I.}~\bibnamefont{Lucas}},
  \bibinfo{author}{\bibfnamefont{K.}~\bibnamefont{Uchida}},
  \bibinfo{author}{\bibfnamefont{P.~A.} \bibnamefont{Algarabel}},
  \bibinfo{author}{\bibfnamefont{L.}~\bibnamefont{Morell{\'{o}}n}},
  \bibinfo{author}{\bibfnamefont{E.}~\bibnamefont{Saitoh}}, \bibnamefont{and}
  \bibinfo{author}{\bibfnamefont{M.~R.} \bibnamefont{Ibarra}},
  \bibinfo{journal}{Physical Review B} \textbf{\bibinfo{volume}{90}},
  \bibinfo{pages}{054422} (\bibinfo{year}{2014}).

\bibitem[{\citenamefont{Wu et~al.}(2010)\citenamefont{Wu, Abid, Chun, Ramos,
  Mryasov, and Shvets}}]{Wu2010}
\bibinfo{author}{\bibfnamefont{H.-C.} \bibnamefont{Wu}},
  \bibinfo{author}{\bibfnamefont{M.}~\bibnamefont{Abid}},
  \bibinfo{author}{\bibfnamefont{B.~S.} \bibnamefont{Chun}},
  \bibinfo{author}{\bibfnamefont{R.}~\bibnamefont{Ramos}},
  \bibinfo{author}{\bibfnamefont{O.~N.} \bibnamefont{Mryasov}},
  \bibnamefont{and} \bibinfo{author}{\bibfnamefont{I.~V.}
  \bibnamefont{Shvets}}, \bibinfo{journal}{Nano Lett.}
  \textbf{\bibinfo{volume}{10}}, \bibinfo{pages}{1132} (\bibinfo{year}{2010}).

\bibitem[{\citenamefont{Wu et~al.}(2012)\citenamefont{Wu, Ramos, Sofin, Liao,
  Abid, and Shvets}}]{Wu2012a}
\bibinfo{author}{\bibfnamefont{H.-C.} \bibnamefont{Wu}},
  \bibinfo{author}{\bibfnamefont{R.}~\bibnamefont{Ramos}},
  \bibinfo{author}{\bibfnamefont{R.~G.~S.} \bibnamefont{Sofin}},
  \bibinfo{author}{\bibfnamefont{Z.-M.} \bibnamefont{Liao}},
  \bibinfo{author}{\bibfnamefont{M.}~\bibnamefont{Abid}}, \bibnamefont{and}
  \bibinfo{author}{\bibfnamefont{I.~V.} \bibnamefont{Shvets}},
  \bibinfo{journal}{Appl. Phys. Lett.} \textbf{\bibinfo{volume}{101}},
  \bibinfo{pages}{052402} (\bibinfo{year}{2012}).

\bibitem[{\citenamefont{Hoffmann}(2013)}]{Hoffmann2013}
\bibinfo{author}{\bibfnamefont{A.}~\bibnamefont{Hoffmann}},
  \bibinfo{journal}{IEEE Trans. Magn.} \textbf{\bibinfo{volume}{49}},
  \bibinfo{pages}{5172} (\bibinfo{year}{2013}).

\bibitem[{\citenamefont{Sinova et~al.}(2015)\citenamefont{Sinova, Valenzuela,
  Wunderlich, Back, and Jungwirth}}]{Sinova2015}
\bibinfo{author}{\bibfnamefont{J.}~\bibnamefont{Sinova}},
  \bibinfo{author}{\bibfnamefont{S.~O.} \bibnamefont{Valenzuela}},
  \bibinfo{author}{\bibfnamefont{J.}~\bibnamefont{Wunderlich}},
  \bibinfo{author}{\bibfnamefont{C.}~\bibnamefont{Back}}, \bibnamefont{and}
  \bibinfo{author}{\bibfnamefont{T.}~\bibnamefont{Jungwirth}},
  \bibinfo{journal}{Rev. Mod. Phys.} \textbf{\bibinfo{volume}{87}},
  \bibinfo{pages}{1213} (\bibinfo{year}{2015}).

\bibitem[{\citenamefont{Orna et~al.}(2010)\citenamefont{Orna, Algarabel,
  Morell{\'{o}}n, Pardo, de~Teresa, {L{\'{o}}pez Ant{\'{o}}n}, Bartolom{\'{e}},
  Garc{\'{i}}a, Bartolom{\'{e}}, Cezar et~al.}}]{Orna2010}
\bibinfo{author}{\bibfnamefont{J.}~\bibnamefont{Orna}},
  \bibinfo{author}{\bibfnamefont{P.~A.} \bibnamefont{Algarabel}},
  \bibinfo{author}{\bibfnamefont{L.}~\bibnamefont{Morell{\'{o}}n}},
  \bibinfo{author}{\bibfnamefont{J.~A.} \bibnamefont{Pardo}},
  \bibinfo{author}{\bibfnamefont{J.~M.} \bibnamefont{de~Teresa}},
  \bibinfo{author}{\bibfnamefont{R.}~\bibnamefont{{L{\'{o}}pez Ant{\'{o}}n}}},
  \bibinfo{author}{\bibfnamefont{F.}~\bibnamefont{Bartolom{\'{e}}}},
  \bibinfo{author}{\bibfnamefont{L.~M.} \bibnamefont{Garc{\'{i}}a}},
  \bibinfo{author}{\bibfnamefont{J.}~\bibnamefont{Bartolom{\'{e}}}},
  \bibinfo{author}{\bibfnamefont{J.~C.} \bibnamefont{Cezar}},
  \bibnamefont{et~al.}, \bibinfo{journal}{Phys. Rev. B}
  \textbf{\bibinfo{volume}{81}}, \bibinfo{pages}{144420}
  (\bibinfo{year}{2010}).

\bibitem[{\citenamefont{Uchida et~al.}(2010{\natexlab{b}})\citenamefont{Uchida,
  Adachi, Ota, Nakayama, Maekawa, and Saitoh}}]{Uchida2010i}
\bibinfo{author}{\bibfnamefont{K.}~\bibnamefont{Uchida}},
  \bibinfo{author}{\bibfnamefont{H.}~\bibnamefont{Adachi}},
  \bibinfo{author}{\bibfnamefont{T.}~\bibnamefont{Ota}},
  \bibinfo{author}{\bibfnamefont{H.}~\bibnamefont{Nakayama}},
  \bibinfo{author}{\bibfnamefont{S.}~\bibnamefont{Maekawa}}, \bibnamefont{and}
  \bibinfo{author}{\bibfnamefont{E.}~\bibnamefont{Saitoh}},
  \bibinfo{journal}{Appl. Phys. Lett.} \textbf{\bibinfo{volume}{97}},
  \bibinfo{pages}{172505} (\bibinfo{year}{2010}{\natexlab{b}}).

\bibitem[{\citenamefont{Ramos et~al.}(2013)\citenamefont{Ramos, Kikkawa,
  Uchida, Adachi, Lucas, Aguirre, Algarabel, Morell{\'{o}}n, Maekawa, Saitoh et~al.
  }}]{Ramos2013}
\bibinfo{author}{\bibfnamefont{R.}~\bibnamefont{Ramos}},
  \bibinfo{author}{\bibfnamefont{T.}~\bibnamefont{Kikkawa}},
  \bibinfo{author}{\bibfnamefont{K.}~\bibnamefont{Uchida}},
  \bibinfo{author}{\bibfnamefont{H.}~\bibnamefont{Adachi}},
  \bibinfo{author}{\bibfnamefont{I.}~\bibnamefont{Lucas}},
  \bibinfo{author}{\bibfnamefont{M.~H.} \bibnamefont{Aguirre}},
  \bibinfo{author}{\bibfnamefont{P.}~\bibnamefont{Algarabel}},
  \bibinfo{author}{\bibfnamefont{L.}~\bibnamefont{Morell{\'{o}}n}},
  \bibinfo{author}{\bibfnamefont{S.}~\bibnamefont{Maekawa}},
  \bibinfo{author}{\bibfnamefont{E.}~\bibnamefont{Saitoh}},
  \bibnamefont{et~al.}, \bibinfo{journal}{Appl. Phys. Lett.}
  \textbf{\bibinfo{volume}{102}}, \bibinfo{pages}{072413}
  (\bibinfo{year}{2013}).

\bibitem[{\citenamefont{Snyder and Toberer}(2008)}]{Snyder2008a}
\bibinfo{author}{\bibfnamefont{G.~J.} \bibnamefont{Snyder}} \bibnamefont{and}
  \bibinfo{author}{\bibfnamefont{E.~S.} \bibnamefont{Toberer}},
  \bibinfo{journal}{Nat. Mat.} \textbf{\bibinfo{volume}{7}},
  \bibinfo{pages}{105} (\bibinfo{year}{2008}).

\bibitem[{\citenamefont{Uchida et~al.}(2014)\citenamefont{Uchida, Ishida,
  Kikkawa, Kirihara, Murakami, and Saitoh}}]{Uchida2014a}
\bibinfo{author}{\bibfnamefont{K.}~\bibnamefont{Uchida}},
  \bibinfo{author}{\bibfnamefont{M.}~\bibnamefont{Ishida}},
  \bibinfo{author}{\bibfnamefont{T.}~\bibnamefont{Kikkawa}},
  \bibinfo{author}{\bibfnamefont{A.}~\bibnamefont{Kirihara}},
  \bibinfo{author}{\bibfnamefont{T.}~\bibnamefont{Murakami}}, \bibnamefont{and}
  \bibinfo{author}{\bibfnamefont{E.}~\bibnamefont{Saitoh}},
  \bibinfo{journal}{J. Phys.: Condens. Matter} \textbf{\bibinfo{volume}{26}},
  \bibinfo{pages}{343202} (\bibinfo{year}{2014}).

\bibitem[{\citenamefont{Sakuraba et~al.}(2013)\citenamefont{Sakuraba, Hasegawa,
  Mizuguchi, Kubota, Mizukami, Miyazaki, and Takanashi}}]{Sakuraba2013}
\bibinfo{author}{\bibfnamefont{Y.}~\bibnamefont{Sakuraba}},
  \bibinfo{author}{\bibfnamefont{K.}~\bibnamefont{Hasegawa}},
  \bibinfo{author}{\bibfnamefont{M.}~\bibnamefont{Mizuguchi}},
  \bibinfo{author}{\bibfnamefont{T.}~\bibnamefont{Kubota}},
  \bibinfo{author}{\bibfnamefont{S.}~\bibnamefont{Mizukami}},
  \bibinfo{author}{\bibfnamefont{T.}~\bibnamefont{Miyazaki}}, \bibnamefont{and}
  \bibinfo{author}{\bibfnamefont{K.}~\bibnamefont{Takanashi}},
  \bibinfo{journal}{Appl. Phys. Express} \textbf{\bibinfo{volume}{6}},
  \bibinfo{pages}{033003} (\bibinfo{year}{2013}).

\bibitem[{\citenamefont{Kikkawa et~al.}(2013)\citenamefont{Kikkawa, Uchida,
  Daimon, Shiomi, Adachi, Qiu, Hou, Jin, Maekawa, and
  Saitoh}}]{Kikkawa_PRB_2013}
\bibinfo{author}{\bibfnamefont{T.}~\bibnamefont{Kikkawa}},
  \bibinfo{author}{\bibfnamefont{K.}~\bibnamefont{Uchida}},
  \bibinfo{author}{\bibfnamefont{S.}~\bibnamefont{Daimon}},
  \bibinfo{author}{\bibfnamefont{Y.}~\bibnamefont{Shiomi}},
  \bibinfo{author}{\bibfnamefont{H.}~\bibnamefont{Adachi}},
  \bibinfo{author}{\bibfnamefont{Z.}~\bibnamefont{Qiu}},
  \bibinfo{author}{\bibfnamefont{D.}~\bibnamefont{Hou}},
  \bibinfo{author}{\bibfnamefont{X.-F.} \bibnamefont{Jin}},
  \bibinfo{author}{\bibfnamefont{S.}~\bibnamefont{Maekawa}}, \bibnamefont{and}
  \bibinfo{author}{\bibfnamefont{E.}~\bibnamefont{Saitoh}},
  \bibinfo{journal}{Phys. Rev. B} \textbf{\bibinfo{volume}{88}},
  \bibinfo{pages}{214403} (\bibinfo{year}{2013}).

\bibitem[{\citenamefont{Lee et~al.}(2015)\citenamefont{Lee, Kim, {Yeon Lee},
  Kim, Lee, Lee, Jeong, Lee, Song, Sohn et~al.}}]{Lee2015}
\bibinfo{author}{\bibfnamefont{K.-D.} \bibnamefont{Lee}},
  \bibinfo{author}{\bibfnamefont{D.-J.} \bibnamefont{Kim}},
  \bibinfo{author}{\bibfnamefont{H.}~\bibnamefont{{Yeon Lee}}},
  \bibinfo{author}{\bibfnamefont{S.-H.} \bibnamefont{Kim}},
  \bibinfo{author}{\bibfnamefont{J.-H.} \bibnamefont{Lee}},
  \bibinfo{author}{\bibfnamefont{K.-M.} \bibnamefont{Lee}},
  \bibinfo{author}{\bibfnamefont{J.-R.} \bibnamefont{Jeong}},
  \bibinfo{author}{\bibfnamefont{K.-S.} \bibnamefont{Lee}},
  \bibinfo{author}{\bibfnamefont{H.-S.} \bibnamefont{Song}},
  \bibinfo{author}{\bibfnamefont{J.-W.} \bibnamefont{Sohn}},
  \bibnamefont{et~al.}, \bibinfo{journal}{Sci. Rep.}
  \textbf{\bibinfo{volume}{5}}, \bibinfo{pages}{10249} (\bibinfo{year}{2015}).

\bibitem[{\citenamefont{Uchida et~al.}(2015)\citenamefont{Uchida, Kikkawa,
  Seki, Oyake, Shiomi, Takanashi, and Saitoh}}]{Uchida_2015}
\bibinfo{author}{\bibfnamefont{K.}~\bibnamefont{Uchida}},
  \bibinfo{author}{\bibfnamefont{T.}~\bibnamefont{Kikkawa}},
  \bibinfo{author}{\bibfnamefont{T.}~\bibnamefont{Seki}},
  \bibinfo{author}{\bibfnamefont{T.}~\bibnamefont{Oyake}},
  \bibinfo{author}{\bibfnamefont{J.}~\bibnamefont{Shiomi}},
  \bibinfo{author}{\bibfnamefont{K.}~\bibnamefont{Takanashi}},
  \bibnamefont{and} \bibinfo{author}{\bibfnamefont{E.}~\bibnamefont{Saitoh}},
  \bibinfo{journal}{Phys. Rev. B} \textbf{\bibinfo{volume}{92}},
  \bibinfo{pages}{094414} (\bibinfo{year}{2015}).

\bibitem[{\citenamefont{Shiomi et~al.}(2015)\citenamefont{Shiomi, Handa,
  Kikkawa, and Saitoh}}]{Shiomi2015}
\bibinfo{author}{\bibfnamefont{Y.}~\bibnamefont{Shiomi}},
  \bibinfo{author}{\bibfnamefont{Y.}~\bibnamefont{Handa}},
  \bibinfo{author}{\bibfnamefont{T.}~\bibnamefont{Kikkawa}}, \bibnamefont{and}
  \bibinfo{author}{\bibfnamefont{E.}~\bibnamefont{Saitoh}},
  \bibinfo{journal}{Appl. Phys. Lett.} \textbf{\bibinfo{volume}{106}},
  \bibinfo{pages}{232403} (\bibinfo{year}{2015}).

\end{thebibliography}

\newpage

\newpage

\newpage

\newpage

\end{document}